\documentclass[preprintnumbers,twocolumn,amsmath,amssymb]{revtex4}

\usepackage{amsmath}
\usepackage{amsfonts}
\usepackage{amssymb}
\usepackage{verbatim}
\usepackage{graphicx}
\usepackage{color}

\newcommand{\beq}{\begin{equation}}
\newcommand{\eeq}{\end{equation}}
\newcommand{\bqa}{\begin{eqnarray}}
\newcommand{\eqa}{\end{eqnarray}}

\newcommand{\bra}[1]{ \langle{#1} |}
\newcommand{\ket}[1]{ |{#1} \rangle}

\definecolor{ngreen}{rgb}{0.2,0.6,0.2}

\definecolor{golden}{rgb}{0.8,0.6,0.1}

\definecolor{purp}{rgb}{0.8,0.1,0.8}

\definecolor{orange}{rgb}{0.9,0.3,0}
\definecolor{mar}{rgb}{0.6,0.1,0.1}

\begin{document}

\title{Quantum State Effusion}

\author{Howard M. Wiseman}

\address{Centre for Quantum Computation and Communication Technology (Australian Research Council), \\ Centre for Quantum Dynamics, Griffith University, Brisbane, QLD 4111, Australia}

\begin{abstract}
I give a scientific perspective, with a personal emphasis, on the seminal 1992 paper, {\em The quantum-state diffusion model applied to open systems}, by Gisin and Percival.
\end{abstract}


\maketitle

1992 was an exciting year for physicists working on open quantum systems. No fewer than 
five groups~\cite{DalCasMol92,TeiMah92,DumZolRit92,GisPer92b,TiaCar92} 
independently introduced the idea that a Markovian open quantum system, such as a laser-driven atom, 
evolving deterministically as a mixed state because of coupling to its environment, 
could be fruitfully modelled as, and perhaps understood as really being in, a stochastically 
evolving pure state. Four of these groups modelled this stochastic behaviour 
as quantum jumps, corresponding (it was suggested) to the emission of a photon by the atom, 
or to its detection. But the paper by Nicolas Gisin and Ian 
Percival~\cite{GisPer92b} postulated something different, 
which they called {\em quantum state diffusion} (QSD). 

Gisin and Percival showed that, for any master equation of the 
Lindblad form~\cite{Lin76}, describing the mixed state $\rho_t$ of an open quantum system, 
one can write down a continuous, stochastic, nonlinear, differential equation for a pure state 
$\ket{\psi_t}$, which, in the mean, reproduces the master equation solution: 
$\rho_t = {\rm M}[\ket{\psi_t}\bra{\psi_t}]$, 
provided this condition holds at $t=0$ 
(e.g.~by having $\rho_0=\ket{\psi_0}\bra{\psi_0}$). 
Their equation involves one complex white noise term for each Lindblad operator 
in the master equation. Now, given a master equation, there 
is no unique way to define the set of Lindblad operators, or even to uniquely separate the Hamiltonian 
term from the Lindblad term(s). But, remarkably, as Gisin and Percival showed, their 
QSD equation is invariant under any choice of how to express the master equation. 
This property sets it apart from the quantum jump equations of Refs.~\cite{DalCasMol92,DumZolRit92,TiaCar92}. 

Like the jump models of Ref.~\cite{DalCasMol92,DumZolRit92,TiaCar92} 
(though unlike that of Ref.~\cite{TeiMah92}), the QSD equation is potentially 
useful for efficiently simulating the dynamics of large open quantum systems. Indeed,
it was adapted into a practical tool for simulating electronic wave-packet dynamics in 
atoms and molecules by Percival and co-workers~\cite{SteAlbPer95}.  
And, beyond being merely useful, it could reveal aspects of the dynamics that were 
invisible in the solution of the master equation. It was 
Spiller and Ralph~\cite{SpiRal94} who applied the QSD equation to study the quantum 
analogue of a chaotic dissipative classical system, enabling them to calculate, 
for the first time, a quantum Poincar\'e map exhibiting many of 
the fine features of the classical attractor.  Building on this, similar quantum 
diffusion equations (see below) were applied to quantitatively evaluate signatures of 
quantum chaos 
such as Lyapunov exponents~\cite{BhaHabJac00}.  
But do the individual trajectories in QSD, whether chaotic or not, 
have any physical meaning? This is where I enter the story.

1992 was a particularly exciting year for a particular physicist (me) beginning a PhD under the supervision 
of Gerard Milburn. It began in January with a research lecture by Howard Carmichael, on the 
topic of quantum trajectories. This was my first introduction to the quantum jump theory 
he published in Ref.~\cite{TiaCar92} and the quantum diffusion theory he soon after 
published~\cite{Car93b}. The quantum jumps, he showed, corresponded to photon detections, 
and his quantum diffusion to homodyne detection, achieved by interfering the emitted field with a 
strong laser prior to detection. But this quantum diffusion theory was 
not quite the same as Gisin and Percival's QSD --- it involved real (not complex) noise, and 
was not invariant under different representations of the same master equation. 
This same equation, 
with the same (though less concretized) 
interpretation, had in fact earlier been derived in the mathematical physics 
community~\cite{BarBel91}. 

In May 1992, Gerard  
showed me a preprint of Gisin and Percival's paper which had appeared in his pigeonhole (yes, 1992 
was a different world). I immediately wanted to know --- did it correspond to some quantum optical 
detection scheme? In June, Gerard sent me an idea by airmail (see preceding parenthetical comment) 
from 
Aspen: heterodyne detection. This is very similar to homodyne detection but uses a 
detection laser far-detuned from the system resonance. Soon, I had shown that Gisin and Percival's 
QSD could be derived using Carmichael's quantum trajectory theory, adapted to heterodyne detection~\cite{WisMil93c}. That is, the QSD equation, its invariance properties notwithstanding,  is 
just one of infinitely many quantum trajectory equations, each corresponding to 
monitoring the system's environment in a different way. Indeed, Percival soon shifted emphasis away 
from QSD as pertaining to a system interacting with an environment, 
and towards interpreting the equations as describing ``primary state diffusion''~\cite{Per94} --- a new and fundamental irreversible physical process (perhaps related to gravity~\cite{Per95}), 
in the tradition of earlier work on the foundations of quantum mechanics~\cite{GRW86,Pea86,Dio88,Gis89}. 

Gisin and Percival's work on QSD in Ref.~\cite{GisPer92b} has, however, continued to influence physicists in the field of open quantum systems. Di\'{o}si and I gave a complete parameterisation of quantum diffusion equations, 
and showed that there are whole families 
that have the same invariance properties as the QSD equation~\cite{WisDio01}. 
Meanwhile, it was shown by Strunz, Di\'{o}si, and Gisin that QSD could be generalized to non-Markovian open quantum systems~\cite{StrDioGis99}. Gambetta and I subsequently
showed that, just as in the Markovian case, QSD is not unique in this regard: non-Markovian quantum trajectories with real noise, reducing to those for homodyne trajectories in the Markovian limit, could also be found~\cite{GamWis02}. 
Indeed, we could understand non-Markovian QSD in the very general context 
of system states `conditioned' on the values of environmental hidden variables~\cite{GamWis03}. 

Returning to Markovian QSD, although Ref.~\cite{WisMil93c} answered, for me, the question as to its 
operational meaning (it describes conditioning on a stochastic heterodyne measurement record, not 
 `conditioning' on observer-independent environmental 
 variables), 
 an important question remained unanswered: can we {\em prove} this experimentally? 
 That is, can we prove that an open quantum system, such as a driven atom, is {\em not} always 
 (that is,  heedless of any monitoring of its environment we might perform) undergoing
QSD? More generally, 
can we prove experimentally that {\em no} observer-independent pure-state dynamical model describes 
the state of such a system 
\footnote{This amounts to testing quantum trajectory theory, in which entanglement between the system 
 and its environment is crucial,  against all rival theories such that the local system is still 
 described by quantum mechanics but is always in a (perhaps hidden) pure state.}?
  Wishing to answer that question led me 
  to Schr\"odinger's notion of steering~\cite{SchSteer35}.  I, with co-workers, formalized this 
 as disproving a hybrid (local quantum state on one side, local hidden variable on the other) model of 
 correlations~\cite{WisJonDoh07} and, finally, 
 applied it to an open quantum system and its environment~\cite{WisGam12,DarWis14}. 
Our conclusion in Refs.~\cite{WisGam12,DarWis14} was that it indeed should be possible, albeit challenging, to do an experiment of this sort. Gisin, meanwhile, has recently found 
QSD trajectories 
for open quantum systems relevant to discussions of free-will and the nature of time~\cite{Gis16}. 
 Thus I am sure that many ramifications will continue to grow  
 from Gisin and Percival's seminal idea in 1992. 
   

\bibliography{GP-perspective-arxiv.bbl}

\end{document}